\documentstyle[12pt]{article}
\def\mathbf{\vec}

\def\ca{\c{c}\~{a}}

\begin{document}
\baselineskip=15pt

\centerline {\LARGE Implications of a New Effective Chiral Meson Lagrangian}
\vspace{1cm}
\centerline {\large Alexander A. Osipov\footnote{On leave from the 
            Laboratory of Nuclear Problems, JINR, 141980 Dubna, Russia}, 
            Marcos Sampaio and Brigitte Hiller}
\vspace{.5cm}
\centerline {\it Centro de F\'{\i}sica Te\'{o}rica, Departamento de
             F\'{\i}sica}
\centerline {\it da Universidade de Coimbra, 3004-516 Coimbra, Portugal}
\vspace{1cm}

\begin{abstract}
Based on a recently derived effective chiral meson Lagrangian from the 
extended $SU(2)\otimes SU(2)$ Nambu -- Jona-Lasinio (ENJL) model, in
the linear realization of chiral symmetry, we extract to leading order 
in the $1/N_c$ expansion all associated relevant three-point functions 
$\rho \rightarrow \pi\pi$, $\sigma \rightarrow \pi\pi$, $a_1
\rightarrow \rho\pi$, $a_1 \rightarrow \sigma\pi$, as well as the amplitude
for $\pi\pi$ scattering. We discuss the formal differences of these amplitudes
as compared with those derived in the literature and calculate the associated 
decay widths and scattering parameters. The differences have two origins: 
i) new terms, which are proportional to the current quark mass and arise 
from taking the correct NJL vacuum from the first steps in a proper-time  
expansion, are present in the Lagrangian; ii) an implemented chiral
covariant treatment of the diagonalization in the pseudoscalar - 
axialvector sector induces new couplings between three or more mesonic fields. 
Both effects have been derived from the chiral Ward Takahashi identities, 
which are fully taken into account at each order of the proper-time expansion.
\end{abstract}

\section{Introduction}

The Nambu -- Jona-Lasinio (NJL) model \cite{Nambu:1961} and its several 
extensions (see e.g. \cite{Ebert:1983}-\cite{Bernard:1997}) have been vastly 
studied as effective models of the strong interaction, based on the chiral 
dynamics of four-quark interactions. By incorporating the main symmetries of 
QCD and being reminiscent of the effective four-fermion interaction for QCD, 
obtained after eliminating the gluonic degrees of freedom \cite{'t Hooft:1976},
the NJL model is a useful playground for simulating relevant features of 
low-energy hadron physics. Its innumerous applications range from the 
calculation of the low-lying meson spectra, meson couplings, decay and 
scattering amplitudes, diquark physics and extensions to the baryonic
sector, to modelling of finite density and temperature effects on
chiral properties of hadrons. 

In the present paper we focus on the implications of a recently derived   
$SU(2)\otimes SU(2)$ effective chiral Lagrangian \cite{Osipov1:2000}, 
\cite{Osipov2:2000}, on the low-lying hadron phenomenology. The Lagrangian 
has been constructed on the basis of the ENJL model by using the Schwinger 
proper-time representation for the modulus of the one-loop quark determinant 
\cite{Schwinger:1951}, \cite{DeWitt:1965}, and the following long wavelength 
expansion of its heat kernel. This semiclassical WKB expansion of the ENJL
action is implemented by polynomial counterterms, which result from requiring 
that the symmetry breaking pattern of the fermionic Lagrangian in the presence
of the explicit chiral symmetry breaking term be equivalent to the one of the 
bosonized effective Lagrangian \cite{Osipov2:2000}, \cite{Osipov:2001}.
As a consequence of these symmetry requirements we have shown how relevant and
previously not considered current quark mass terms appear in the local action 
of the chiral mesonic fields. These terms allow to account for the correct 
vacuum already at the first steps of the proper-time expansion and lead to 
a resummation in the current quark mass. Furthermore, in the case of the 
linear realization of chiral symmetry, the Lagrangian contains new meson 
couplings which derive from a modified diagonalization of the axialvector - 
pseudoscalar interaction, which we have shown to be necessary in order to 
preserve the chiral transformation properties of the vector mesons. We 
consider worthwhile understanding the consequences of such new structures on 
relevant amplitudes and scattering processes. As we shall show, the effects of
the new current quark mass terms will be manifest in all considered 
amplitudes, and may appear both explicitly and implicitly through the coupling
parameters. As for the covariant diagonalization, we shall observe the 
following: 1) the $\rho\pi\pi$ coupling becomes a three-derivative type, 
contrary to the one-derivative type obtained in the usual linear approaches. 
This is important, since the latter violates chiral symmetry 
\cite{Gasser:1984}. On the mass shell one recovers the one-derivative 
structure. 2)The $a_1\pi\sigma$ coupling 
acquires also new three-derivative type of couplings. On the mass shell it 
reduces in form to the known results of \cite{Osipov:1992}, 
\cite{Bijnens:1993}. 3)The amplitudes, $\sigma\pi\pi$ and $a_1\rho\pi$ are not
altered by the proposed covariant diagonalization. 4) The contact term with  
four pion fields gets modified with extra two- and four-derivatives in the 
fields. One expects therefore that when off-shell processes are at work, such 
as in form factors or $\rho$-exchange in $\pi\pi$ scattering, the related
amplitudes are affected correspondingly. We shall show, however, that in 
the case of $\pi\pi$ scattering, one recovers old results (up to current quark mass terms), regardless of using the covariant diagonalization.

We work in the leading $1/N_c$ approximation, that is, to fermion one-loop 
level. The bosonized Lagrangian is correspondingly treated to tree level order
in the meson couplings. Furthermore we sum the proper-time series up to the 
third Seeley-DeWitt coefficient, which amounts to keeping, out of the full 
momentum dependence of the n-point functions, only the quadratic and 
logarithmic divergent contributions. There are several reasons to stop at this
order in the heat kernel expansion. First, once this is done, the masses 
and coupling constants of the chiral fields are completely fixed in a way 
which guarantees that the first and the second Weinberg sum rules are 
automatically satisfied. 
Second, truncating the proper-time series at this 
order is substantiated by the results of \cite{De Rafael:1998}, where an 
infinite number of local counterterm operators were added to the ENJL 
Lagrangian, with couplings fixed, such that the corresponding Adler function 
exhibited the properties of the ``lowest meson dominance" approximation to 
large-$N_c$ QCD. For the vector and axial-vector two-point functions this 
requirement was tantamount to removing the non-confining terms and guarantees
their correct matching to the QCD short-distance behaviour. Third, one might 
expect, except for amplitudes which are finite previous to regularization, the
divergent contributions to dominate over the finite ones. 
    
The paper is structured as follows. In the second section we set up the 
notation and situate the problem, by giving a short review of the results 
obtained in \cite{Osipov1:2000}, \cite{Osipov2:2000}. In section 3 we derive 
the amplitudes $\rho \rightarrow \pi\pi$, $\sigma \rightarrow \pi\pi$, $a_1 
\rightarrow \rho\pi$ and $a_1 \rightarrow \sigma\pi$ and discuss the 
differences with respect to similar amplitudes obtained from other models
based on ENJL-type Lagrangians. In section 4 we derive the amplitude 
for $\pi\pi$ scattering. In section 5 we present the numerical results.
We conclude with a summary and outlook.

\section{The Lagrangian: current quark mass terms and covariant 
         diagonalization}

The starting point is the effective quark Lagrangian of strong interactions 
which is invariant under a global colour $SU(N_c)$ symmetry
\begin{eqnarray}
\label{enjl}
  {\cal L}&=&\bar{q}(i\gamma^\mu\partial_\mu -m_c)q
            +\frac{G_S}{2}[(\bar{q}q)^2+(\bar{q}i\gamma_5\tau_i q)^2]
            \nonumber\\
          &-&\frac{G_V}{2}[(\bar{q}\gamma^\mu\tau_i q)^2
            +(\bar{q}\gamma^\mu\gamma_5\tau_i q)^2].
\end{eqnarray}
Here $q$ is a flavour doublet of Dirac spinors for quark fields $\bar q=(\bar 
u, \bar d)$. Summation over the colour indices is implicit. We use the standard
notation for the isospin Pauli matrices $\tau_i$. The current quark mass matrix
$m_c=\mbox{diag}(\hat m_u,\hat m_d)$ is chosen in such a way that 
$\hat m_u=\hat m_d=\hat m$. Without
this term the Lagrangian (\ref{enjl}) would be invariant under global chiral 
$SU(2)\otimes SU(2)$ symmetry. The coupling constants $G_S$ and $G_V$ have
dimensions $(\mbox{Length})^2$ and can be fixed from the meson mass spectrum.
The transformation law for the quark fields is the following
\begin{equation}
\label{quark}
   \delta q=i(\alpha +\gamma_5\beta )q, \quad
   \delta\bar{q}=-i\bar{q}(\alpha -\gamma_5\beta )
\end{equation}
where parameters of global infinitesimal chiral transformations are chosen as
$\alpha =\alpha_i\tau_i, \ \ \beta =\beta_i\tau_i$. 
Under infinitesimal chiral transformations the Lagrangian ${\cal L}$ 
exhibits therefore the following explicit symmetry breaking pattern
\begin{equation}
\label{sb}
   \delta {\cal L}=-2i\hat{m}(\bar{q}\gamma_5\beta q).
\end{equation}
which is to be kept intact at each stage of calculations (here we are not 
considering the anomalous sector). The chiral effective Lagrangian which we 
obtain \cite{Osipov2:2000} from (\ref{enjl}) as result of the heat kernel 
expansion up to and including the third order Seeley -- DeWitt coefficient  
and taking into account the symmetry requirements has the following form
in the spontaneously broken phase
\begin{eqnarray}
\label{fleff}
   {\cal L}_{\mbox{eff}}
   &=&\frac{v_{\mu i}^2+a_{\mu i}^2}{2G_V}-\frac{\hat{m}(\sigma^2 
     +\vec{\pi}^2)}{2(m-\hat{m})G_S}
     -\frac{N_cJ_1}{8\pi^2}\left[\frac{1}{6}\mbox{tr}(v_{\mu\nu}^2
     +a_{\mu\nu}^2)\right.\nonumber \\
   &-&\left.\frac{1}{2}\mbox{tr}\left((\nabla_\mu\pi )^2
     +(\nabla_\mu\sigma )^2\right)
     +\left(\sigma^2+2(m-\hat{m})\sigma +\pi^2_i\right)^2\right]
\end{eqnarray}
where the trace is to be taken in isospin space. Here we have used the notation
\begin{equation}
\label{vmn}
   v_{\mu\nu}=\partial_\mu v_\nu -\partial_\nu v_\mu -i[v_\mu ,v_\nu ]
             -i[a_\mu ,a_\nu ],
\end{equation}     
\begin{equation}
\label{amn}
   a_{\mu\nu}=\partial_\mu a_\nu -\partial_\nu a_\mu -i[a_\mu ,v_\nu ]
             -i[v_\mu ,a_\nu ],
\end{equation}     
\begin{equation}
   \nabla_\mu\sigma =\partial_\mu\sigma -i[v_\mu ,\sigma ]+\{a_\mu ,\pi\},
\end{equation}    
\begin{equation}
   \nabla_\mu\pi =\partial_\mu\pi -i[v_\mu ,\pi ]-\{a_\mu ,\sigma +m-\hat{m}\}.
\end{equation}
with $v_\mu=v_{\mu i}\tau_i, a_\mu=a_{\mu i}\tau_i, \sigma,\pi=\pi_i\tau_i$ 
designating the vector isovector, axialvector isovector, scalar isoscalar and 
pseudoscalar isovector fields respectively, and $m$ is the constituent 
quark mass. In terms of these fields the infinitesimal chiral transformation 
laws read
\begin{equation}
\label{st}
   \delta\sigma =-\{\beta, \pi\}, \quad
   \delta\pi =i[\alpha, \pi ]+2(\sigma +m-\hat{m})\beta ,
\end{equation}
\begin{equation}
\label{transva}
   \delta v_\mu =i[\alpha, v_\mu ]+i[\beta , a_\mu ], \quad
   \delta a_\mu =i[\alpha, a_\mu ]+i[\beta , v_\mu ].
\end{equation}
The variation of the second term of (\ref{fleff}) yields 
the symmetry breaking pattern of the Lagrangian in terms of the collective 
fields, which is the equivalent of eq.(\ref{sb}) in terms of the fermionic 
variables 
\begin{equation}
\label{sbm}
   \delta {\cal L}=-\frac{2\hat{m}}{G_S}(\beta_i\pi_i )
                  =\delta {\cal L}_{\mbox{eff}}.
\end{equation}
All other terms in (\ref{fleff}) are chiral invariant.
The function $J_1$ appearing in (\ref{fleff}) is one of the set of integrals 
$J_n$ emerging in the heat kernel expansion \cite{Osipov2:2000}, 
\begin{equation}
J_n(m^2,\Lambda^2)=\int^\infty_0\frac{dT}{T^{2-n}}e^{-Tm^2}\rho (T,\Lambda^2),
         \quad n=0,1,2...
\end{equation}
In the explicit evaluation of these integrals we use as regulating kernel
the Pauli-Villars cutoff \cite{Pauli:1949} with two subtractions
\begin{equation}
\label{cc}
      \rho (T, \Lambda^2)=1-(1+T\Lambda^2)e^{-T\Lambda^2}.
\end{equation}
Prior to regularization the $J_1$ integral is logarithmically divergent.
The other characteristic divergence of the ENJL model at one-loop order 
is the quadratic one, given by $J_0$, which has been traded by the gap 
equation in writing down (\ref{fleff})
\begin{equation}
\label{gap}
  \frac{m-\hat{m}}{mG_S}=\frac{N_cJ_0}{2\pi^2},
\end{equation}
to establish the real vacuum of the spontaneously broken phase. As it stands 
the effective Lagrangian (\ref{fleff}) still requires a diagonalization of 
the pseudoscalar-axialvector fields appearing in the quadratic forms for the 
covariant derivatives. We have shown in \cite{Osipov2:2000} that the simplest 
replacement of variables which fulfills the linear transformation property 
(\ref{transva}) not only for old variables, $v_\mu , a_\mu$, but also
for new ones, $v'_\mu , a'_\mu$, is
\begin{eqnarray}
\label{ared}
  a_\mu&=&a'_\mu +\frac{\kappa}{2}\left(\{\sigma +m-\hat{m}, \partial_\mu\pi\}
         -\{\pi ,\partial_\mu\sigma\}\right),\nonumber \\
  v_\mu&=&v'_\mu +\frac{i\kappa}{2}\left([\sigma , \partial_\mu\sigma ]
         +[\pi ,\partial_\mu\pi ]\right).
\end{eqnarray}
For the case at hand the commutator $[\sigma , \partial_\mu\sigma ]=0$.
These redefinitions involve new terms that are bilinear in the fields 
and induce changes at the level of couplings with three or more fields, as 
compared to the non-covariant diagonalizations that have widely been used 
previously in the linear chiral symmetry versions of the ENJL model. The 
replacement (\ref{ared}) is identical to the field redefinition considered 
in \cite{Bijnens:1993} for the case of non-linear realization of chiral 
symmetry. The constant $\kappa$ is fixed by the requirement that the 
bilinear part of the effective Lagrangian becomes diagonal in the fields 
$\pi , a'_\mu$. We find in this way that
\begin{equation}
\label{kappa}
   \frac{1}{2\kappa}=\left(m-\hat{m}\right)^2+\frac{\pi^2}{N_cJ_1G_V}.
\end{equation}

The physical meson fields are obtained as usual by bringing the kinetic 
terms to their standard form. For the vector fields one has:

\begin{equation}
\label{renorm1}
   v'_\mu =\sqrt{\frac{6\pi^2}{N_cJ_1}}v_\mu^{(ph)}\equiv
           \frac{g_{\rho}}{2}v_\mu^{(ph)}, 
   \quad   a'_\mu =\frac{g_{\rho}}{2}a_\mu^{(ph)}.
\end{equation}
Then we have
\begin{equation}
\label{mass1}
   m^2_\rho =\frac{6\pi^2}{N_cJ_1G_V}, \quad m^2_a=m^2_\rho +6(m-\hat{m})^2.
\end{equation}
In particular it implies the relations 
\begin{equation}
\label{Z}
   g_A=1-\frac{6(m-\hat{m})^2}{m^2_a}=\frac{m^2_\rho}{m^2_a},
   \quad \kappa =\frac{3}{m^2_a}.
\end{equation}
We also have to redefine the spin-0 fields
\begin{equation}
\label{renorm0}
  \sigma =\sqrt{\frac{4\pi^2}{N_cJ_1}}\sigma^{(ph)}\equiv
          g_{\sigma}\sigma^{(ph)}, \quad \pi =g_\pi\pi^{(ph)}, 
          \quad g_\pi =\frac{g_\sigma}{\sqrt{g_A}} .
\end{equation}
The mass formulae for spin-0 fields are
\begin{equation}
\label{mass0}
   m^2_\pi =\frac{\hat{m}g^2_\pi}{(m-\hat{m})G_S}, \quad 
   m^2_\sigma =g_Am^2_\pi +4(m-\hat{m})^2.
\end{equation}
As compared with previous calculations in \cite{Ebert:1986,Bijnens:1993} our 
mass formulae have a different dependence on the current quark mass. 

Let us also point out that after the field redefinitions the symmetry breaking 
part takes the form \cite{Gasior:1969}
\begin{equation}
\label{sbf}
   \delta {\cal L}_{\mbox{eff}}=-2m^2_\pi f_\pi \beta_i\pi_i^{(ph)}
\end{equation}
which leads to the well known PCAC relation for the divergence of the quark
axial-vector current 
\begin{equation}
\label{pcac}
   \partial_\mu\vec{J}^\mu_A=2f_\pi m^2_\pi \vec{\pi}^{(ph)}.
\end{equation}
We used the relation
\begin{equation}
\label{gpi}
   g_\pi =\frac{m-\hat{m}}{f_\pi}
\end{equation}
to get (\ref{sbf}).

\section{Three-meson vertices}

{\bf 1. The $\sigma\pi\pi$ interaction}
\vspace{0.5cm}

After using the field redefinitions (\ref{ared}) in the effective Lagrangian
(\ref{fleff}) and collecting all terms involving one scalar and two 
pseudoscalar fields one gets
\begin{eqnarray}
\label{spp}
   {\cal L}_{\sigma\pi\pi}
   &=&-2\frac{g_{\sigma}}{g_A}(m-\hat{m})\sigma\left\{\left[1-
     \frac{m_{\pi}^2(1-g_A)}{2(m-\hat{m})^2}\right]\vec{\pi}^2
     \right.\nonumber \\
   &+&\left.\frac{1-g_A^2}{2(m-\hat{m})^2}(\partial_\mu\vec{\pi})^2\right\}
\end{eqnarray} 
where use has been made of the mass relation for the scalar field
eq. (\ref{mass0}), the field renormalizations (\ref{renorm0}) (here and 
henceforth we drop the index (ph) on the physical fields),and the
mixing parameter $\kappa$, eq. (\ref{kappa}). One obtains for the decay 
$\sigma(q)\rightarrow\pi_a(p) \pi_b(p')$ (a,b are isospin indices) 
the amplitude
\begin{equation}
\label{mspp}
   {\cal M}_{\sigma\pi\pi}(p,p')=\frac{1}{4}Tr\{\tau_a,\tau_b\}
                                 f_{\sigma\pi\pi}(p,p').
\end{equation}
\begin{equation}
\label{fsppo}
   f_{\sigma\pi\pi}(p,p')=4\frac{g_{\sigma}}{g_A}(m-\hat{m})
   \left\{1-\frac{1-g_A}{2(m-\hat{m})^2}\left[m_\pi^2+pp'(1+g_A)
   \right]\right\}.
\end{equation}
On the mass-shell $m_\sigma^2=2(m_\pi^2+pp')$ and using (\ref{mass0}) 
for the $\sigma$ mass one obtains
\begin{equation}
\label{fspp}
   f_{\sigma\pi\pi}=4g_{\sigma}(m-\hat{m})\left\{g_A
    +\frac{m_{\pi}^2(1-g_A)^2}{4(m-\hat{m})^2}\right\}.
\end{equation}
This amplitude differs from previously calculated ones by the current 
quark mass terms. For instance in \cite{Bernard:1996}: keeping only the 
logarithmically divergent integrals at zero squared momentum, we find a 
correspondence to the considered order of the present heat kernel expansion, 
after the substitutions ((lhs) are the notations of \cite{Bernard:1996} and 
(rhs) the present)
\begin{equation}
\label{sub}
      m\rightarrow m-\hat{m}, \qquad 
      \delta=1-\frac{6m^2}{m_a^2}\rightarrow g_A,
\end{equation} 
which lead to eq. (\ref{fspp}). The decay width is obtained in the 
standard way
\begin{equation}
\label{ds}
    \Gamma_{\sigma\pi\pi} = \frac{3f^2_{\sigma\pi\pi}}{8\pi m_\sigma^2}
                            \sqrt{(m_\sigma^2-4m_\pi^2)}
\end{equation}

\vspace{0.5cm}
{\bf 2. The $\rho\pi\pi$ interaction}
\vspace{0.5cm}

Again by collecting all terms involving the $v_{\mu}$ meson field and two
pseudoscalar fields, after the redefinition (\ref{ared}), one obtains the
interaction Lagrangian
\begin{eqnarray}
\label{rpp}
   {\cal L}_{\rho\pi\pi}
   &=&ig_{\rho}g_\pi^2\mbox{tr}\left\{\frac{\kappa}{8G_V} v_\mu
     [\pi,\partial_\mu \pi] \right. \nonumber\\
   &-&\frac{N_cJ_1}{16\pi^2}\left[\frac{\kappa}{3}
     \left(1-\kappa(m-\hat m)^2\right)
     \tilde v_{\mu\nu}[\partial_\mu \pi, \partial_\nu \pi]
     \right. \nonumber\\
   &-&\left.\left. (2\kappa(m-\hat m)^2-1)\partial_\mu \pi
     [v_\mu,\pi]\right]\right\}.
\end{eqnarray}
with $\tilde v_{\mu\nu}$ denoting the derivative terms of (\ref{vmn}). 
The terms containing $\kappa$ which are not multiplying quark mass factors
stem from the proposed field bilinears in the redefinition (\ref{ared}) 
for the vector fields and were therefore absent in previous analyses. Also 
all $\hat m$ terms are new. Using eqs. (\ref{kappa}) and (\ref{Z}) one can 
recast the interaction in the form
\begin{equation}
\label{rpp1}
   {\cal L}_{\rho\pi\pi}=-i\frac{g_{\rho}(1+g_A)}{8m_\rho^2}\mbox{tr}
   \left(\tilde v_{\mu\nu}[\partial_\mu{\pi},\partial_\nu{\pi}]\right).
\end{equation}
The interaction is of three-derivative type, as opposed to the usual 
one-derivative coupling. This is a consequence of the chiral covariant 
diagonalization. On the mass-shell one obtains, after partial integration 
in the action and discarding total derivatives
\begin{equation}
\label{rpp0}
   {\cal L}_{\rho\pi\pi}=
   -i\frac{g_{\rho}}{8}(1+g_A)\mbox{tr}(v_{\mu}[\pi,\partial_{\mu}\pi]).
\end{equation}
The Lagrangian becomes on-shell equivalent in form to the standard 
expression, for the non-linear as well as linear cases, see e.g. 
\cite{Bijnens:1993} and \cite{Bernard:1996}.
In order to make these comparisons one should again keep only the 
logarithmically divergent contributions in the cases considered in 
\cite{Bijnens:1993}, \cite{Bernard:1996}, to be compatible with
the order of the heat kernel expansion considered in the present approach.
Starting from the Lagrangian ${\cal L}_{\rho\pi\pi}$ of \cite{Bijnens:1993}
\begin{eqnarray}
\label{bijrho}
   {\cal L}_{\rho\pi\pi}&=&\frac{-ig_V}{2\sqrt{2}}
   tr(V_{\mu\nu}[\xi^\mu,\xi^\nu]),
   \nonumber\\
   \xi_\mu&=&i(\xi^+ \partial_\mu \xi - \xi\partial_\mu \xi^+)
   \rightarrow \frac{1}{f_\pi}\partial_\mu \pi +... ,\nonumber\\
   \xi&=&\mbox{exp}\left(\frac{-i}{2f_\pi}\lambda_i \phi_i\right),
\end{eqnarray}
one has the following correspondence between the notation of 
\cite{Bijnens:1993} (lhs) and the present one (rhs):
\begin{eqnarray}
\label{corr}
g_V&=&\frac{N_c}{48\pi^2 f_V}(1-g_A^2)\Gamma (0,x)\rightarrow
      \frac{1-g_A^2}{2g_\rho}\nonumber\\
f_V^2&=&\frac{N_c\Gamma (0,x)}{24\pi^2}\rightarrow\frac{1}{g_\rho^2}\nonumber\\
     &&\Gamma(0,x)\rightarrow J_1\nonumber\\
     &&g_A\rightarrow \frac{m_\rho^2}{m_{a_1}^2}
\end{eqnarray}
where $x=m^2/\Lambda_\chi^2$. In the expression for $g_V$ of
\cite{Bijnens:1993} we have already dropped a term proportional to 
$\Gamma(1,x)$, which would correspond to a $J_2$ integral in our notation 
and therefore be of higher order than the one considered in the heat kernel 
expansion of the present work.
Note that, although there is a formal equivalence to the standard result, 
hidden information stemming from the current quark mass terms is carried 
by the expressions relating $m_\rho$ to $m_{a_1}$ and by $g_A$, eqs. 
(\ref{mass1}) and (\ref{Z}).

The amplitude for the process $\rho_{\mu}^a (q)\rightarrow\pi^b(p)\pi^c 
(p')$ is 
\begin{equation}
\label{mrpp0}
   {\cal M}_{\rho\pi\pi}(p,p')=\frac{1}{4}Tr(\tau_a[\tau_b,\tau_c])
   (p-p')_{\mu}\epsilon^{\mu}(q)f_{\rho\pi\pi}
\end{equation}
where $\epsilon^{\mu}(q)$ is the polarization of the vector particle 
and
\begin{equation}
\label{frpp}
   f_{\rho\pi\pi} = \frac{g_\rho}{2}(1+g_A)
\end{equation}
for on-shell particles. The decay width is then 
\begin{equation}
\label{drho}
    \Gamma_{\rho\pi\pi} = \frac{|\vec{p_c}|^3}{6\pi m_\rho^2}
                            f_{\rho\pi\pi}^2
\end{equation}  
with $\vec{p_c}$ being the center of mass momentum of the pions, 
$|\vec{p_c}|=\sqrt{(m^2_\rho -4m^2_\pi )}/2$.

\vspace{0.5cm}
{\bf 3. The $a_1\rho\pi$ and $a_1\sigma\pi$  interactions}
\vspace{0.5cm}

These processes are interesting in relation to the branching ratio
Br$(a_1\rightarrow\pi(\pi\pi)_s)$. According to Weinberg \cite{Weinberg:1990},
chiral symmetry arguments lead to the prediction 
Br$(a_1\rightarrow\pi(\pi\pi)_s)=10-15\%$, in conflict with the value
quoted in Particle Data until 1996 \cite{PD:1996}. The main source of 
the $(\pi\pi)_s$ pairs is the scalar particle decay and the main
decay channel for $a_1$ is $a_1\rightarrow\rho\pi$. The ratio of these 
two decay modes for the $a_1$ should then represent a reasonable estimate 
of the branching ratio. 
 
The Lagrangians for the couplings $a_1\rho\pi$ and $a_1\sigma\pi$ are obtained
in a similar way as in the previous cases
\begin{equation}
\label{arp}
   {\cal L}_{a_1\rho\pi}=if_\pi\frac{g_{\rho}^2}{4g_A}\mbox{tr}
   \left\{\frac{\kappa}{3}\left(a_\mu[\partial_{\nu}\pi,
   \tilde v_{\mu\nu}]+v_\mu [\partial_{\nu}\pi,
   \tilde a_{\mu\nu}]\right) + a_\mu[v_\mu,\pi]\right\},
\end{equation}
\begin{equation}
\label{asp}
   {\cal L}_{a_1\sigma\pi}=g_{\rho}\frac{1-g_A}{\sqrt{g_A}}\mbox{tr}
   \left\{\sigma a_\mu \partial_\mu \pi + \frac{1}{12(m-\hat{m})^2} 
   \tilde a_{\mu\nu} (\partial_{\mu}\pi \partial_{\nu}\sigma -
   \partial_{\nu}\pi \partial_{\mu}\sigma)\right\},
\end{equation}
with $\tilde a_{\mu\nu}$ the derivative terms contained in (\ref{amn}).
Contrary to the case of ${\cal L}_{\rho\pi\pi}$ interaction, the 
terms bilinear in the fields in eq. (\ref{ared}) do not contribute to
${\cal L}_{a_1\rho\pi}$. The terms proportional to $\kappa$ stem only from
the linear combination $\kappa(m-\hat m)\partial_\mu \pi$ of the shift in the
$a_\mu$ field. On the mass-shell the interaction Lagrangians reduce to
\begin{equation}
\label{arp0}
   {\cal L}_{a_1\rho\pi}
   =if_\pi\frac{g_{\rho}^2}{4}\mbox{tr}(a_{\mu}[v_\mu,\pi])
\end{equation}
which coincides in form with the results in 
\cite{Bijnens:1993}, \cite{Osipov:1992}, \cite{Prades:1994}  
\noindent and to
\begin{equation}
\label{asp0}
   {\cal L}_{a_1\sigma\pi}=-g_{\rho}\sqrt{g_A}\mbox{tr}
   (\sigma a_{\mu} \partial_{\mu}\pi )
\end{equation}
which corresponds to the result of \cite{Bijnens:1993},
\cite{Osipov:1992}. Let us note, however, that the couplings $g_A$ and 
$f_\pi$ depend, in the present approach, on extra current quark mass
terms, eqs. (\ref{Z}) and (\ref{gpi}).

The decay amplitudes for the processes 
  $a_{1 \mu}^a (q) \rightarrow \rho_\nu^b (p) \pi^c (p')$ and 
  $a_{1 \mu}^a (q) \rightarrow \sigma (p) \pi^b (p')$ on-shell are 
\begin{equation}
\label{marp0}
   {\cal M}_{a_1\rho\pi}(p,p')=-\frac{i}{4}Tr(\tau_a[\tau_b,\tau_c]) 
   \epsilon_{\mu}(q)\epsilon_{\nu}^*(p) f_{a_1\rho\pi}^{\mu\nu},
\end{equation}
\begin{equation}
\label{masp0}
   {\cal M}_{a_1\sigma\pi}(p,p')=\frac{i}{4}Tr\{\tau_a,\tau_b\}
   \epsilon_{\mu}(q) p'^\mu f_{a_1\sigma\pi},
\end{equation}
with
\begin{equation}
\label{farp0}
   f_{a_1\rho\pi}^{\mu\nu}=f_\pi g_{\rho}^2 g^{\mu\nu}
\end{equation}
\begin{equation}
\label{fasp0}
   f_{a_1\sigma\pi}=2 g_\rho \sqrt{g_A}
\end{equation}
Finally the decay widths are calculated to be 
\begin{equation}
\label{dar}
    \Gamma_{a_1\rho\pi} = \frac{f_\pi^2g_\rho^4}{12\pi m_a^3}
    \left(2+\frac{(qp)^2}{m_a^2 m_\rho^2}\right)\sqrt{(qp)^2-m_a^2m_\rho^2}
\end{equation}  
with $2qp=m_a^2+m_\rho^2-m_\pi^2$ and
\begin{equation}
\label{das}
    \Gamma_{a_1\sigma\pi}=\frac{m_a}{192\pi}|f_{a_1\sigma\pi}|^2
    \left\{\left[1-\left(\frac{m_\sigma+m_\pi}{m_a}\right)^2\right]
    \left[1-\left(\frac{m_\sigma-m_\pi}{m_a}\right)^2\right]
    \right\}^{\frac{3}{2}}.
\end{equation}  

\section{$\pi\pi$ scattering}

The scattering amplitude $T_{ab;cd}$ for the process
$\pi^a(p_1)+\pi^b(p_2)\rightarrow\pi^c(p_3)+\pi^d(p_4)$ has the well-known 
isotopic structure
\begin{equation}
\label{pipi}
   T_{ab;cd}(s,t,u)=\delta_{ab}\delta_{cd}A(s,t,u)
                   +\delta_{ac}\delta_{bd}A(t,s,u)
                   +\delta_{ad}\delta_{cb}A(u,t,s).
\end{equation}  
Here, the standard Mandelstam variables for two-particle elastic scattering, 
$s, t$ and $u$, are defined by 
\begin{equation}
\label{mand}
   s=(p_1+p_2)^2, \quad t=(p_1-p_2)^2, \quad u=(p_1-p_4)^2.
\end{equation}  
Amplitudes with definite isospin $(I)$, $T^I$, are then
\begin{eqnarray}
\label{iso}
    T^0(s,t,u) 
    &=& A(t,s,u)+A(u,t,s)+A(s,t,u),\nonumber\\ 
    T^1(s,t,u) 
    &=& A(t,s,u)-A(u,t,s),\nonumber\\ 
    T^2(s,t,u) 
    &=& A(t,s,u)+A(u,t,s).
\end{eqnarray}  
After the redefinitions (\ref{ared}) the Lagrangian (\ref{fleff}) contributes 
with scalar and $\rho$ meson exchange as well as with a contact term to the 
scattering amplitude. 
First we evaluate the scalar exchange amplitude $A_{\sigma}(s,t,u)$ 
using the interaction Lagrangian (\ref{spp}) for the $\sigma\pi\pi$ vertex  
\begin{eqnarray}
\label{sce}
    A_{\sigma}(s,t,u)
    &=&\frac{16g_\pi^2(m-\hat{m})^2}{g_A(m_{\sigma}^2-s)}
       \left\{\left[1-\frac{m_\pi^2(1-g_A)}{2(m-\hat{m})^2}\right]^2
       \right.\nonumber \\
    &-&\left[1-\frac{m_\pi^2(1-g_A)}{2(m-\hat{m})^2}\right]
       \frac{(1-g_A^2)(s-2m_\pi^2)}{(m-\hat{m})^2}\nonumber \\   
    &+&\left.\frac{(1-g_A^2)^2(s-2m_\pi^2)^2}{16(m-\hat{m})^4}\right\}.   
\end{eqnarray}
The scalar propagator is expanded up to a desired order in $(m-\hat{m})^{-2}$,
\begin{eqnarray}
\label{sexp}
    \frac{1}{m_\sigma^2-s}    
    &=&\frac{1}{4(m-\hat{m})^2}\left[1+\frac{g_Am_\pi^2-s}
      {4(m-\hat{m})^2}\right]^{-1}\nonumber\\
    &=& \frac{1}{4(m-\hat{m})^2}\left\{1-\frac{g_Am_\pi^2-s}{4(m-\hat{m})^2}+
        \frac{(g_Am_\pi^2-s)^2}{16(m-\hat{m})^4} +... \right\}
\end{eqnarray}
leading to  
\begin{equation}
\label{sexpf}
    A_{\sigma}(s,t,u)=4\frac{g_\pi^2}{g_A}+\frac{g_A}{f_\pi^2}
    \left[s\left(2-\frac{1}{g_A^2}\right)
    -m_\pi^2\left(4-\frac{3}{g_A}\right)\right]+...
\end{equation}

Next we obtain the $\rho$ exchange amplitude $A_{\rho}(s,t,u)$ 
using the interaction Lagrangian (\ref{rpp}) for the $\rho\pi\pi$ vertex.  
The $\rho$-propagator has the conventional form
\begin{equation}
\label{rhp}
    \Delta_{\mu\nu}^{ab}(x_1-x_2)=-i\delta_{ab}\int \frac{d^4 k}{(2\pi)^4} 
    \left(\frac{k_\mu k_\nu}{m_\rho^2}-g_{\mu\nu}\right)
    \frac{e^{-ik(x_1-x_2)}}{(m_\rho^2-k^2-i\epsilon)}
\end{equation}
\noindent and the amplitude reads
\begin{equation}
\label{re}
    A_{\rho}(s,t,u)=\frac{g_\rho^2}{m_\rho^4}
    (1+g_A)^2\left\{\frac{t^2(s-u)}{m_\rho^2-t}
    +\frac{u^2(s-t)}{m_\rho^2-u}\right\}.
\end{equation}
This amplitude starts at ${\cal O}(p^6)$ in chiral counting and therefore 
does not contribute to the Weinberg result. It can be compared to the $\rho$ 
meson exchange contribution to the $\pi\pi$ scattering amplitude derived by 
Gasser and Leutwyler \cite{Gasser:1984}, in spite of the fact that their 
$\rho$ meson has origin in an antisymmetric tensor field. This is because  
in the evaluation of the respective $S$-matrix element, ${\cal M}_{4\pi}$, 
with $\rho$-exchange   
\begin{eqnarray}
\label{smre}
    && {\cal M}_{4\pi}=\frac{-ig_\rho^2 (1+g_A^2)}{128m_\rho^2}         
       (\delta_{il}\delta_{jm}-\delta_{im}\delta_{jl})
       \int d^4 x_1 d^4 x_2 \nonumber\\
    && \langle \pi^c(p_3)\pi^d(p_4)|\partial_{\mu}\pi_i(x_1)
       \partial_{\nu}\pi_j(x_1)\partial_{\alpha}\pi_l(x_2)
       \partial_{\beta}\pi_m(x_2){\cal U}_{\mu\nu\alpha\beta}
       |\pi^a(p_1)\pi^b(p_2)\rangle
\end{eqnarray}
one encounters the term ${\cal U}_{\mu\nu\alpha\beta}$:
\begin{equation}
\label{smu}
  {\cal U}_{\mu\nu\alpha\beta} = \int \frac{d^4 k}{(2\pi)^4} 
         \frac{e^{-ik(x_1-x_2)}}{(m_\rho^2-k^2)}
         \left[-k_\mu k_\alpha g_{\nu\beta} + k_\nu k_\alpha g_{\mu\beta}
         -k_\nu k_\beta g_{\mu\alpha}+k_\mu k_\beta g_{\nu\alpha}\right]
\end{equation}
which is identical to a piece of the Green's function for the $\rho$ meson in 
\cite{Gasser:1984} and leads to the contribution proportional to 
$(m_\rho^2-k^2)^{-1}$ for the corresponding $\pi\pi$ scattering amplitude.

Finally we evaluate the contact term, by collecting all terms with four 
pseudoscalar fields          
\begin{eqnarray}
\label{ct0}
   {\cal L}_{c}
   &=&g_\pi^4\mbox{tr}\left\{
      \frac{-\kappa^2}{16G_V}[\pi, \partial_\mu\pi]^2
      +\frac{N_cJ_1}{8\pi^2}\left[\frac{\kappa^2}{6}(1-\kappa (m-\hat m)^2)^2
      [\partial_\mu\pi, \partial_\nu\pi]^2-\frac{\pi^4}{2} 
      \right. \right. \nonumber\\
   &+&\frac{\kappa^2}{2}(m-\hat m)^2
      \{\partial_\mu\pi, \pi\}^2+
      \frac{\kappa}{2}\partial_\mu\pi [[\pi,\partial_\mu\pi ],\pi ]
      \nonumber\\
   &+&\left.\left. \kappa^2(m-\hat m)^2\partial_\mu 
      \pi [\pi, [\pi, \partial_\mu\pi ]] \right] \right\}.
\end{eqnarray}
All terms containing powers of $\kappa$ as factors (except the ones 
multiplying quark masses) derive from the covariant diagonalization
and were not present in previous schemes. After some rearrangement one 
can write the contact term as
\begin{eqnarray}
\label{ct}
    {\cal L}_{c}
    &=&-\frac{g_\pi^2}{2g_A}(\vec{\pi}^2)^2+\frac{(1-g_A)}{2g_A f_\pi^2}
      \left[(\vec{\pi}\partial_\mu \vec{\pi})^2-g_A\vec{\pi}^2
      (\partial_\mu \vec{\pi})^2\right]\nonumber\\
    &-&(1-g_A^2)(1+g_A)
      \frac{(\partial_\mu \vec{\pi}\times\partial_\nu \vec{\pi})^2}
       {4f_\pi^2m_\rho^2},
\end{eqnarray}
which leads to the contact amplitude 
\begin{eqnarray}
\label{camp}
    A_c(s,t,u)
    &=&-4\frac{g_\pi^2}{g_A} + \frac{(1-g_A)}{f_\pi^2 g_A}\left[s+
      2g_A(s-2m_\pi^2)\right]\nonumber\\
    &+&\frac{(1-g_A^2)(1+g_A)}{4f_\pi^2m_\rho^2}\left[(s-t)u+(s-u)t\right].
\end{eqnarray}

Now we obtain the complete amplitude $A(s,t,u)$ for $\pi\pi$ scattering, by
assembling the scalar and vector exchange amplitudes and contact terms. 
We give here explicitly the result to $p^4$-th order   
\begin{eqnarray}
\label{fou}
        A(s,t,u)
        &=&\frac{1}{f_\pi^2}(s-m_\pi^2) \nonumber\\
        &+&\frac{1}{24g_Af_\pi^2 (m-\hat m)^2}\Big\{(1-g_A^2)^2
        \Big[(s-t)u+(s-u)t\Big]\nonumber\\
        &+&6g_A^2\Big[(s-2m_\pi^2)g_A+m_\pi^2\Big]^2\Big\}+{\cal O}(p^6). 
\end{eqnarray}
This result can be compared with the one obtained using the non-covariant 
diagonalization in \cite{Bernard:1996} after keeping there only the 
logarithmic divergent contributions at zero squared momentum, again to 
relate to the order of heat kernel expansion considered in the present work.
To order $p^2$ we obtain the Weinberg result \cite{Weinberg:1967}.
In fact we find that for any $p^2$ order of $A(s,t,u)$ one recovers the 
previous result, except for the current quark mass terms (i.e. if one puts 
everywhere in (\ref{fou}) $m-\hat m \rightarrow m$). It turns out that in 
the case of the usual non-covariant diagonalization and induced linear 
derivative $\rho$ coupling to the pions, a judicious combination of the 
chiral non-covariant terms emerging in the vector channel and the contact 
term simulates the correct structure of the contact term obtained in the case 
of the covariant diagonalization, up to $p^4$ order (the vector exchange 
only starts at $p^6$ order). Starting from ${\cal O}(p^6)$ the vector
exchange term coincides in the two approaches and the contact term does not 
contribute in both cases, at the considerd order of the heat kernel expansion.

In the next section we analyze the numerical effects due to the present heat
kernel expansion on $\pi\pi$ threshold parameters, as compared to the studies 
where the full momentum expansion of Feynman amplitudes was considered 
\cite{Bernard:1996}. We do not expect large deviations, since, at least to 
the order of the heat kernel expansion considered here, 
the $\pi\pi$ amplitude does not get modified by the covariant diagonalization,
and the current quark mass may not be large enough to make the new terms in
the amplitude numerically significant. However it is worthwhile measuring the 
numerical effects related to the momentum expansions in the two approaches,
since they differ by finite non-vanishing contributions present in 
\cite{Bernard:1996}, due to differences of logarithmic divergent integrals 
of different arguments and other finite terms.

\section{Numerical results}

We start the numerical section by calculating the decay widths of the heavy 
mesons. The four parameters of the model, $G_S$, $G_V$, $\Lambda$ and 
$\hat m$, are obtained by fixing $m_\pi=139$ MeV, $f_\pi=92$ MeV, 
$m_\rho=770$ MeV and the ratio $g_A=m_\rho^2/m_a^2$, which we take at two 
different $m_a$ values. For $g_A=0.5$, in accordance with the choice of 
\cite{Weinberg:1990}, we obtain $m_a=1089$ MeV, $m=314$ MeV, 
$\hat m=1.7$ MeV, $G_S=3.22$ GeV$^{-2}$, $G_V=14.77$ GeV$^{-2}$, 
$\Lambda=1.536$ GeV, and for $g_A=0.374$, which corresponds to the empirical 
$m_a=1260$ MeV, we get $m=408$ MeV, $\hat m=1.4$ MeV, $G_S=3.40$ GeV$^{-2}$, 
$G_V=$18.49 GeV$^{-2}$, $\Lambda=1.544$ GeV. In Table I we display the  
mesonic observables, set $I$ corresponding to the smaller $m$ value
and set $II$ to the large one.
\begin{table}
\begin{center}
\caption[]{
Some meson properties calculated in the present version of the NJL model are 
compared to experimental data \cite{Pdg:2000}. The asterisks indicate 
quantities which served as input to determine the model parameters.}
\begin{tabular}{|c|c|c|c|}
\hline
&&&\\
$[MeV]$& model (set $I$) & model (set $II$)&  experiment \\
&&&\\
\hline
$f_\pi$ & $92^*$   & $92^*$ &$93.3$     \\
$m_\pi$ & $139^*$  & $139^*$  & $139$      \\
$m_\sigma$ & $633$& $818$    & $400 - 1200$ \\
$m_\rho$ & $770^*$& $770^*$    & $770$ \\
$m_{a_1}$ & $1089^*$  &$1260^*$& $1260$ \\
$\Gamma_{\sigma \to \pi\pi}$ & $409$ & $394$    & $600 - 1000$ \\
$\Gamma_{\rho\to \pi\pi}$ & $82$& $86$     & $150$ \\
$\Gamma_{a1\to \rho\pi}$ & $192$ & $420$    & seen; Full $\Gamma=250-600$\\
$\Gamma_{a1\to \sigma\pi}$ & $31$& $23$     & seen  \\
&&\\
\hline
\end{tabular}
\end{center} 
\end{table}
Some comments are in order here. The decay width $\Gamma_{\sigma\pi\pi}$ and 
$\Gamma_{\rho\pi\pi}$ turn out to be smaller than the empirical values by 
roughly a factor two, if one insists on keeping the correct empirical fit 
for $m_\pi$ and $f_\pi$. 
This trend did not change as compared to the calculations in a full momentum 
scheme, for the cases in which the $\rho$-meson is a well-defined bound state 
below the quark-antiquark pair threshold; the $\sigma$-meson in the latter 
case is always slightly embedded in the continuum (with a very small decay 
width in quark-antiquark pairs \cite{Bernard:1997}).   
The results for the branching ratio
\begin{equation}
\label{bran}
    Br[a_1\rightarrow\pi(\pi\pi)_s]\sim 
    \frac{\Gamma_{a_1\rightarrow\sigma\pi}}{
    \Gamma_{a_1\rightarrow\sigma\pi}+\Gamma_{a_1\rightarrow\rho\pi}} 
    \sim 14\%(I); 5\%(II)
\end{equation}
are in fair agreement to the $10-20\%$ obtained by Weinberg.
The rather large change observed in the width $\Gamma_{a_1\rightarrow\rho\pi}$
from set $II$ of parameters to $I$  is mainly dictated by the square root 
term in eq. (\ref{dar}), which is reduced by roughly a factor two 
and the change in the coupling $g_\rho^2$, which gets smaller by $\sim 25\%$.

Next we present in Table II the results for threshold parameters $a_l^I$,
$b_l^I$ from the representation (\ref{fou}) of the $\pi\pi$ scattering
amplitude, as compared to the data of \cite{Froggatt:1977}. The more recent 
analysis of \cite{Truol:2001} yields $a_0^0=0.288\pm 0.012\pm 0.003$ and   
$a_0^2=-0.036\pm 0.009$. Let us note that since we are working at meson tree 
level, the $p^2$ expansion in our case reveals the subjacent quark-antiquark
compositeness of the amplitudes and therefore we do not compare it to the 
meson-loop orders related momentum expansion of CHPT (for recent reviews see 
\cite{Amoros:2000} and \cite{Colangelo:2001}.)  
\begin{table}
\begin{center}
\caption[]{
The calculated $\pi\pi$ scattering lengths and effective ranges are compared 
to Soft Meson Theorems (SMT) \cite{Weinberg:1967} and experimental data 
(taken from \cite{Froggatt:1977}, and \cite{Truol:2001}(see text please).
}
\begin{tabular}{|c|c|c|c|c|c|c|c|}
\hline
&&&&&&&\\
$a_l^I$&${\cal O}(p^4)[I]$&${\cal O}(p^6)[I]$&${\cal O}(p^4)[II]$&
${\cal O}(p^6)[II]$& Full \cite{Bernard:1996}& SMT 
& experiment\cite{Froggatt:1977}\\
&&&&&&&\\
\hline
&&&&&&&\\
$a_0^0$ &$0.166$ & $0.167$ & $0.161$&$0.162$&$0.17$&$0.16$& $0.26\pm 0.05$\\
$b_0^0$ &$0.188$ & $0.192$ &$.178$& $0.179$&$0.19$& $0.18$& $0.25\pm 0.03$\\
$a_0^2$ &$-0.0454$& $-0.0454$ &$-0.0454$&$-0.0454$ &$-0.047$&$-0.0454$
& $-0.028\pm 0.012$\\
$b_0^2$ &$-0.0875$& $-0.0875$ &$-0.0875$&$-0.0875$&$-0.090$ &$-0.089$ & 
$-0.082\pm 0.008$ \\
$a_1^1$ &$0.0336$& $0.0347$ &$0.0338$ &$0.0351$ &$0.038$&$0.030$& $0.038\pm 0.002$ \\
$a_2^0\times 10^4$ &$5.92$& $8.198$ &$5.034$ &$7.476$&$6.9$& & $17\pm 3$ \\
$a_2^2\times 10^4$&$-0.74$&$-1.93$& $-1.96$ &$-3.16$&$-2.5$& & $1.3\pm 3$ \\
&&&&&&&\\
\hline
\end{tabular}
\end{center} 
\end{table}

At $p^2$ order (not shown in the table) the quantities $a_0^0$, $b_0^0$, 
$a_0^2$, $b_0^2$, $a_1^1$ reproduce the soft pion theorem values. 
The observed trend of $\pi\pi$ scattering lengths and effective ranges 
is congruent with the results of the full momentum expansion 
\cite{Bernard:1996} (included in the table, for the larger value of the 
constituent quark mass considered there ($m=390$MeV)). Some deviations 
are observed in the D-wave scattering lengths. We have checked that at 
$p^8$ order there are no significant changes for any of the calculated 
scattering lengths and effective ranges, from which we infer that the 
differences in the higher partial waves are related to the presence of 
finite terms in \cite{Bernard:1996}, not existent in the
heat kernel expansion (see also discussion at the end of previous section). 

In the present calculation, the vector exchange has still a noticeable 
contribution for $a_1^1$ (the scalar exchange and contact contributions 
stabilize at $p^4$ order), as well as in $a_2^0$ and $a_2^2$.  
 
The case of rather small quark mass ($\simeq 200$ MeV) also considered in 
\cite{Bernard:1996} (since it describes better the scalar form factor of 
the pion) was calculated as well 
with the present method, leading to similar conclusions as for the large mass 
case. Nevertheless we do not consider further this case here, 
since the corresponding parameter set yields worse results for the 
heavy meson decays.  

In the light of the numerical results, the present heat kernel expansion 
yields comparable results and trends in the momentum expansion for the 
scattering parameters calculated from Feynman amplitudes with a full 
momentum dependence in the vertices \cite{Bernard:1996}, for the S and 
P waves. Some sizeable effects are observed in the higher partial waves.   

\section{Summary and outlook}

The main concern of this work was to show the formal and numerical 
implications of a recently derived effective $SU(2)\otimes SU(2)$ chiral 
Lagrangian with linear realization of chiral symmetry on mesonic 
observables: mass spectra, strong decays and $\pi\pi$ scattering 
parameters. The considered Lagrangian was constructed on the basis of
the Schwinger-DeWitt proper-time method applied to the ENJL model.
The resulting semiclassical WKB expansion of the ENJL action has been 
done around the correct NJL vacuum state, defined by the corresponding
Schwinger-Dyson equation in the case with explicit chiral symmetry
breaking. We derive that the amplitudes carry the signature of this vacuum: 
the amplitudes get relevant current quark mass corrections, not present in 
previous approaches. Furthermore we also derive that amplitudes with three 
or more fields are affected by the diagonalization in the 
pseudoscalar-axialvector sector which was implemented to correctly describe 
the vector meson chiral transformations for the linear realization of chiral 
symmetry. We have studied in detail the structure of the amplitudes 
$\sigma\pi\pi$, $\rho\pi\pi$, $a_1\rho\pi$ and $a_1\sigma\pi$ as well as 
$\pi\pi$ scattering. The  $\rho\pi\pi$, $a_1\sigma\pi$ amplitudes and the 
contact four pion interaction get modified by the covariant diagonalization 
and all studied amplitudes depend on current quark mass terms.  
On the mass shell we obtain that all studied processes
are not affected by the covariant diagonalization, becoming structurally 
identical (except for current quark mass terms) to the ones obtained in the 
non-linear as well as linear realizations of chiral symmetry, 
at the same order of the heat kernel expansion. 
In this study the current quark mass effects are numerically negligible,
as expected for the SU(2) case. From the formal point of view, however,
the way the new structures appear in the amplitudes hints at possible 
large numerical deviations for the SU(3) case.  
The extension to the SU(3) case is presently under study.  

\vspace{5.0mm}
{\bf Acknowledgements}
\vspace{5.0mm}

This work is supported by grants provided by Funda\ca o para a Ci\^encia e a
Tecnologia, PRAXIS/C/FIS/12247/1998, PESO/P/PRO/15127/1999, 
POCTI/35304/FIS/2000, CERN/P/FIS/40119/2000, PRAXIS XXI/BDP/22016/99 and 
NATO ``Outreach" Cooperation Program.

\baselineskip 12pt plus 2pt minus 2pt

\end{document}